\documentclass[aps,pre,floatfix,superscriptaddress,amsmath,showpacs,twocolumn]{revtex4}
\usepackage{graphicx}

\begin{document}

\title{Statistical regimes of random laser fluctuations}

\author{Stefano Lepri}
\email{stefano.lepri@isc.cnr.it}
\affiliation{Istituto dei Sistemi Complessi, Consiglio Nazionale
delle Ricerche, via Madonna del Piano 10, I-50019 Sesto Fiorentino, Italy}

\author{Stefano Cavalieri}
\affiliation{Dipartimento di Fisica, via G. Sansone 1 I-50019, Sesto
Fiorentino, Italy }
\affiliation{European Laboratory for Non-linear Spectroscopy, via 
N. Carrara 1, I-50019 Sesto Fiorentino, Italy}

\author{Gian-Luca Oppo}
\affiliation{SUPA and Department of Physics,  
University of Strathclyde, 107 Rottenrow, Glasgow, G4 0NG, Scotland, U.K.}

\author{Diederik S. Wiersma}
\affiliation{European Laboratory for Non-linear Spectroscopy, via 
N. Carrara 1, I-50019 Sesto Fiorentino, Italy}
\affiliation{BEC-INFM Center, I-38050 Povo, Trento, Italy}

\date{\today}

\begin{abstract}
Statistical fluctuations of the light emitted from amplifying random 
media are  studied theoretically and numerically. The characteristic scales 
of the diffusive motion of light lead to Gaussian or power-law (L\'evy) 
distributed fluctuations depending on external control parameters. 
In the L\'evy regime, the output pulse is highly irregular 
leading to huge deviations from a mean--field description. Monte Carlo 
simulations of a simplified model which includes the population of the 
medium, demonstrate the two statistical regimes and provide a comparison 
with dynamical rate equations. Different statistics of the fluctuations helps 
to explain recent experimental observations reported in the  
literature.
\end{abstract}
\maketitle

\section{Introduction}
Optical transport in disordered dielectric materials can be described as
a multiple scattering process in which light waves are randomly scattered 
by a large number of separate elements. To first approximation this gives 
rise to a diffusion process. The propagation of light
waves inside disordered dielectric systems shows several analogies with 
electron transport in conducting solids \cite{book} and the transport of 
cold atom gasses \cite{atom_loc}. A particularly interesting situation arises 
when optical gain is added to a random material. In such materials light 
is multiply scattered and also amplified. They can be realized, for instance, 
in the form of a suspension of micro particles with added laser dye or by 
grinding a laser crystal. Optical transport in such systems is described 
by a multiple scattering process with amplification.

If the total gain becomes larger then the losses, fluctuations grow and 
these systems exhibit a lasing threshold. The simplest form of
lasing in random systems is based on diffusive feedback~\cite{Letokhov} 
where a diffusion process traps the light long enough inside the sample to
reach an overall gain larger then the losses. Interference effects do not 
play a role in this form of random lasing. Diffusive random lasing has been 
observed in various random systems with gain, including powders, laser dye 
suspensions, and organic systems~\cite{markushev,migus93,lawandy94andsha94,bahoura02,wiersma01nature}. 
The behavior of such system shows several analogies with a regular laser, 
including gain narrowing, laser spiking and relaxation oscillations
\cite{migus93,wiersma96}. Reports in literature of complex emission spectra from
random lasers containing a collection of spectrally narrow structures
\cite{cao99, cao00b,vardeny_overview} have triggered a debate about the
possibility of lasing of Anderson localized modes in random systems~\cite{cao99}. 
Although Anderson localized modes can, in principle, form very interesting 
lasers resonators in a gain medium~\cite{pradhan94, genack_loc},
there has been no experimental evidence to date of random lasing 
and localization in the same 3D random sample. In general, the observed 
spectra can be understood via a multiple scattering description based on 
the amplification of statistically rare long light paths
that does not require localization or even interference~\cite{Sushil}. These
emission spectra exhibit a strongly chaotic behaviour, related to the
statistical properties of the  intensity above and around the laser threshold.

Theoretical descriptions of light transport in amplifying disordered
media  and random lasing have been based so far on a diffusive 
mechanism~\cite{zyuzin94,john96,wiersma96},  using, for instance, a 
master-equation approach~\cite{Florescu}. To accommodate the existence
of  localization related effects in the diffusion regime, `anomalously
localized states' have been proposed
\cite{altshuler91,mirlin00,patra03,karpov93,skipetrov04,apalkov02}. 
Other attempts to describe random lasing include  Monte Carlo simulations
\cite{Berger,Sushil2}, finite difference time domain  calculations
\cite{Jiang}, and an approach using random matrix
theory~\cite{beenakker9800}.  A common feature of these studies is that
the statistical properties of  a disordered optical system change with
the addition of optical gain.  Random lasers were found to exhibit, for
instance, strong fluctuations of  their laser threshold
\cite{anglos,vandermolen}. It was also proposed that  such systems can
exhibit L\'evy type statistics in the distribution of 
intensities~\cite{Sharma}. 

In this paper we report on a detailed study of the statistical fluctuations of 
the emitted light from random amplifying  media, using both general theoretical
arguments and results from numerical studies. We find that the characteristic
scales  of the diffusive motion of photons lead to Gaussian or power-law
(L\'evy)  distributed fluctuations depending on external parameters. The L\'evy
regime is limited to a specific range of the gain length, and Gaussian
statistics is recovered in the limit of both low and high gain. Monte Carlo
simulations of a simplified model which includes the medium's population, 
and parallel processing of a large number of random walkers,  demonstrate the
two statistical regimes and provide a comparison with dynamical rate
equations. 

In Section~\ref{s:gen} we present some general arguments to explain
the origin of the L\'evy statistics in random amplifying media.  
In addition, we discuss the possibility of observing 
different statistical regimes. To check the validity of the 
proposed general scenario, we define a simple stochastic model
that is suitable for numerical simulations (Section~\ref{s:model}). The 
rate equation corresponding to its mean--field limit are 
introduced in  Section~\ref{s:mf}. The results of Monte Carlo 
simulations are presented in Section~\ref{s:mc} and discussed 
in the context of experimental results in the concluding Section.

\section{Statistics of the emitted light}
\label{s:gen}

Let us consider a sample of optically active material where photons can
propagate and diffuse. Our description is valid in the diffusive regime, hence 
we assume here that $\lambda$ is smaller then the mean free path $\ell$  
($\lambda < \ell$).  The origin of the L\'evy statistics can be understood by
means of the following reasoning. Spontaneously emitted photons are amplified
within the active medium due to stimulated  emission. Their emission energy is
exponentially large in the path length $l$, i.e.
\begin{equation}
I(l) \;=\; I_0\exp(l/\ell_G) \quad,
\label{expI}
\end{equation} 
where we have introduced the gain length $\ell_G$.
On the other hand, the path length in a diffusing medium is 
a random variable with exponential probability distribution
\begin{equation}
p(l) \;=\; \frac{\exp(-l/\langle l\rangle)}{ \langle l\rangle} \quad,
\label{expl}
\end{equation} 
where $\langle l\rangle$ is the average length of the photon path 
within the sample. 
The path length depends on both the geometry of the sample and the 
diffusion constant $D$. A simple estimate of $\langle l \rangle$ can 
be provided by noting that for a diffusive process with diffusion
coefficient $D$, $\langle l\rangle$ is proportional to the mean 
first--passage time yielding \cite{Redner}
\begin{equation}
\langle l\rangle \;=\; \frac{v }{D \Lambda}
\label{lmed}
\end{equation} 
where $v$ is the speed of light in the medium and $\Lambda$ is the  smallest
eigenvalue of the Laplacian in the active domain (with absorbing  boundary
conditions). For instance, $\Lambda=q^2$ with $q=\pi/L$ for an infinite slab 
or a sphere with $L$ being the thickness or the radius, respectively 
\cite{Letokhov}.

The combination of Eqs.~(\ref{expI}) and (\ref{expl}) immediately 
provides that the probability distribution of the emitted intensity 
follows a power--law 
\begin{equation}
p(I) \;=\; \frac{\ell_G }{\langle l\rangle} \, I^{-(1+\alpha)}
\quad, \qquad \alpha \;=\;  \frac{ \ell_G}{\langle l\rangle}\quad .
\label{pow} 
\end{equation}
Obviously the heavy--tail in Eq.~(\ref{pow}) holds asymptotically and 
the distribution should be cut--off below some value $I_0$ .
The properties of the L\'evy distribution (more properly termed 
L\'evy--stable) are well known \cite{Levy}. The most striking one is that 
for $0<\alpha \le 2$ the average $\langle I \rangle$ exists but the 
variance (and all higher--order moments) diverges. This has important
consequences on the statistics of experimental measurements, 
yielding highly irreproducible data with lack of self--averaging
of sample--to--sample fluctuations. On the contrary, for $\alpha>2$
the standard central--limit theorem holds, and fluctuations are 
Gaussian.

The gain length $\ell_G$ is basically controlled by population inversion 
of the active medium. Increasing the latter, $\ell_G$  
and the exponent $\alpha$ (see Eq.~(\ref{pow})) decrease thus enhancing 
the fluctuations. At first glance, one may thus infer that the larger the 
pumping, the stronger the effect. On the other hand, $\ell_G$ is a 
time--dependent quantity that should be determined self--consistently from 
the dynamics. Indeed, above threshold the release of a huge number of photons 
may lead to such a sizeable depletion of the population inversion to force  
$\ell_G$ to increase. It can then be argued that when the depletion is 
large enough, the L\'evy fluctuations may hardly be detectable.

To put the above statements on a more quantitative ground we need to 
estimate the lifetime of the population as created by the 
pumping process. Following Ref.~\cite{Letokhov}, we write the threshold 
condition as 
\begin{equation}
r \;=\; v/\ell_G - D\Lambda \;>\; 0
\label{thr}
\end{equation}
which is interpreted as ``gain larger than losses", the latter being 
caused by the diffusive escape of light from the sample. 
Note that the condition $r=0$ along with Eqs.~(\ref{lmed}) 
and (\ref{pow}) implies $\alpha=1$ at the laser threshold.

For short pump pulses, the time necessary for the intensity to become 
large is of the order of the inverse of the growth rate $r$. When this time 
is smaller than the average residence time within the sample 
$\langle l\rangle/v$, a sizeable amplification occurs on average for each 
spontaneously emitted photon, leading to a strong depletion of the population
inversion. In this case we expect a Gaussian regime where a mean field 
description is valid. The conditions for the L\'evy regime are therefore
$1/r > \langle l\rangle/v $ and $\alpha \le 2 $ and can 
be written as:
\begin{equation} 
\frac{1}{2}\frac{v }{D \Lambda} \;<\; \ell_G \;<\;  
 2\frac{v}{D \Lambda} \quad. 
\label{bounds}
\end{equation} 
Note that the lower bound of the above inequalities correspond to 
$\alpha=1/2$.

\begin{figure}[ht]
\begin{center}
\includegraphics[width=8cm,clip]{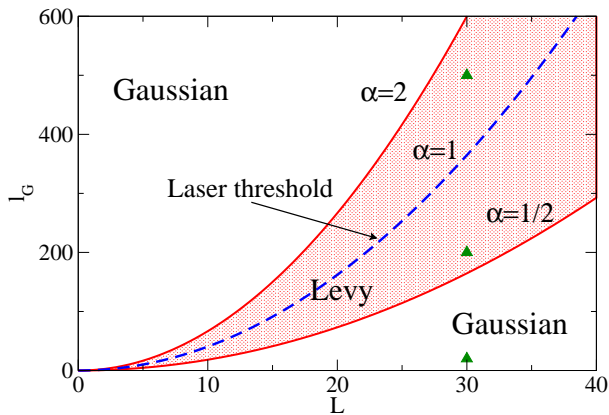}
\caption{(Color online) Different statistical regimes of fluctuations 
of a random laser with a two-dimensional slab geometry of thickness $L$. 
For comparison with the simulations reported below, all quantities 
are expressed in dimensionless units $\ell=1$, $v=1$. The symbols 
correspond to the parameters of Fig.~\ref{f:pulse}.}
\label{f:sog}
\end{center}
\end{figure}

Without losing generality and for later convenience, let us focus on 
the case of a two--dimensional infinite slab of thickness $L$. In 
Fig.~\ref{f:sog} we graphically summarize equation (\ref{bounds}) by drawing a 
diagram in the $(L,\ell_G)$ plane. This representation allows to locate 
three different regions corresponding to different 
statistics. For convenience, the line corresponding to the threshold 
$\alpha=1$ is also drawn. The three regions of statistical interest 
displayed in Fig.~\ref{f:sog} are:

\textit{Subthreshold L\'evy:} weak emission with L\'evy statistics with 
$1<\alpha<2$ (shaded region in Fig.~\ref{f:sog} above the laser threshold 
line);

\textit{Suprathreshold L\'evy:} strong emission with L\'evy statistics 
with $1/2<\alpha\le 1 $
(shaded region in Fig.~\ref{f:sog} below the laser threshold line);

\textit{Gaussian:} $\alpha<1/2$ strong emission with Gaussian statistics,
$\alpha>2$ weak emission with Gaussian fluctuations.  

Note that the first region corresponding to a nonlasing state, should
display anomalous fluctuations as ``precursors" of the transition. It should be
also emphasized that the boundary  between L\'evy and Gaussian statistics is not
expected to correspond  to a sharp transition (as displayed in Fig.~\ref{f:sog})
but rather to a crossover region. 
 
\section{Numerical Model}
\label{s:model}

In order to provide evidence of the theoretical considerations presented  
in Section \ref{s:gen}, we introduce a general, yet easy to simulate, model 
of random lasing. We consider a sample partitioned in cells of linear size 
$\ell$. Specifically, we deal with a portion of a two-dimensional square 
lattice. Thus the center of each cell is identified by the vector index 
${\bf r}=(x,y)$, with $x,y$ integers. In the following we will deal with a 
sample with a slab geometry i.e. $1 \le x\le L$, $1 \le y\le RL$. The total 
number of lattice sites is thus $RL^2$ where $R$ defines the slab aspect 
ratio. Periodic boundary conditions in the $y$ direction are assumed.

Within each cell we have the population $N( {\bf r},t)$ of excitations. 
We consider an hypothetical three--level system with fast decay from the 
lowest laser level. If the population in the latter can be neglected we 
can identify $N$ as the number of atoms in the excited state 
of the lasing transition. 

Isotropic diffusion of light is modeled as a standard random walk along the 
lattice sites. The natural time unit of the dynamics is thus given by 
$\Delta t = \ell/v$.  We choose to describe the diffusion dynamics in terms 
of a set of $M$ walkers each carrying a given number of photons 
$n_1, \ldots n_{M}$. This may be visualized as an ensemble of  ``beams" 
propagating independently throughout the sample. Each of their intensities 
changes in time due to processes of stimulated and spontaneous emission. 
A basic description of those phenomena can be given in terms of a suitable 
master equation \cite{Carma,Florescu} that would require to take into account 
the discrete nature of the variables. To further simplify the model we 
consider that the population and number of photons within each 
cell are so large for the evolution to be well approximated by the 
deterministic equations for their averages. In other words, the rate of 
radiative processes is much larger than that of the diffusive ones and 
a huge number of emissions occurs within the time scale $\Delta t$ \cite{note}. 
With these simplifications $N$ and $n$ can be treated as continuous 
variables. Altogether the model is formulated by the following  
discrete--time dynamics:

\textit{Step 0: pumping -} The active medium is excited homogeneously at the 
initial time i.e. $N( {\bf r},0) = N_0$. The value $N_0$ represents the 
pumping level due to some external source. The initial number of walkers is 
set to $M=0$.

\textit{Step 1: spontaneous emission -} At each time step and for 
every lattice site a spontaneous emission event randomly occurs with 
probability $\gamma N\Delta t$, where $\gamma$ denotes the spontaneous 
emission rate of the single atom. The local population is decreased 
by one:
\begin{equation}
N  \, \longrightarrow \, N-1 \quad,
\end{equation}
and a new walker is started from the corresponding site with 
initial photon number $n=1$. The number of walkers $M$ is increased 
by one accordingly. 

\textit{Step 2: diffusion - } Parallel and asynchronous update of the photons'
positions is performed. Each walker moves with equal probability to one of its
four nearest neighbours. If the boundaries $x=1,L$ of the system are reached,
the walker is emitted and its photon number $n_{out}$ recorded in the output. 
The walker is then removed from the simulation and $M$ is diminished by one. 

\textit{Step 3: stimulated emission -} At each step, the photon numbers $n_i$ 
of each walker and the population are updated deterministically according to 
the following rules: 
\begin{eqnarray}
&& n_i \longrightarrow (1 + \gamma\Delta t \, N)\,n_i \quad, \\   
&& N \longrightarrow (1 - \gamma\Delta t \, n_i)\,N \quad , \nonumber
\end{eqnarray}
where $N$ is the population at the lattice site on which the $i$--th
walker resides.

Stochasticity is thus introduced in the model by both the randomness of 
spontaneous emission events (Step 1) and the diffusive process (Step 2). 
Note that the model in the above formulation does not include
non--radiative decay mechanisms of the population. Furthermore, 
no dependence on the wavelength is, at present, accounted for; in general
$\gamma=\gamma(\lambda)$. 

The initialization described in Step 0 is a crude modelling of the
pulsed pumping employed experimentally. It amounts to considering an
infinitely short excitation during which the samples absorbs $N_0$ photons
from the pump beam. As a further simplification we also assumed that the
excitation is homogeneous throughout the whole sample. More realistic pumping
mechanisms can be easily included in this type of modeling.
More importantly, as we are going to study the time dependence 
of the emission, this type of scheme applies to the case in which
the time separation between subsequent pump pulses is much larger than the 
duration of the emitted pulse (i.e. no repumping effects are present).

Steps 1-3 are repeated up to a preassigned maximum number of iterations.
The sum of all the photon numbers of walkers flowing out of the medium 
at each time step is recorded. The resulting time series is binned on a 
time window of duration $T_W$ to reconstruct the output pulse as it 
would be measured by an external photoncounter. This insures that 
each point is a sum over a large number of events and makes 
the comparison with ensemble-averaged results of the preceding
Section sensible.

It should be emphasized that, although each walker evolves independently
from all the others, they all interact with the same population distribution, 
which, in turn, determines the photon number distributions. In spite
of its simplicity, the model therore describes these two quantities in a 
self-consistent way.

For convenience, we chose to work henceforth in dimensionless units such 
that $v=1$, $\ell=1$ (and thus $\Delta t=1$). The only independent parameters 
are then $\gamma$, the initial population $N_0$ (i.e. the pumping level) 
and the slab sizes $L$, $RL$.

\section{Mean--field equations}
\label{s:mf}

Before discussing the simulation of the stochastic model it is convenient
to present some results on its mean-field limit.
When both the population and photon number are large we expect  
the dynamics to be described by the rate equations for the 
macroscopic averages. This means that, up to relatively small fluctuations, 
the individual realization of the stochastic process should 
follow the solution of \cite{Letokhov,Florescu}
\begin{eqnarray}
&&\dot{N} = -\gamma N(I + 1) \\
&&\dot{I} = D\Delta I + 
\gamma N  (I + 1)
\label{rateq}
\end{eqnarray}
where $I(\textbf{r},t)$ is the number of photons in each cell, $\Delta$ 
denotes the two--dimensional discrete Laplacian and $D=1/4$ in our case. 

For simplicity, let us consider the case of a laterally infinite slab 
($R \to \infty$) in which both $N$ and $I$ depend on the $y$ coordinate, 
only.  Absorbing boundary conditions are imposed, $I(0,t)=I(L+1,t)=0$. 
The integration is started from the same initial conditions of the 
stochastic simulations, namely $N(x,0)=N_0$, $I(x,0) =0$.

As a first remark, we note that the threshold condition (\ref{thr}) applies 
to (\ref{rateq}) upon identifying 
\begin{equation}
\gamma N_0 = \frac{1}{\ell_G}\quad, \qquad q=\frac{\pi}{L+1}.
\label{lg}
\end{equation}
We can thus define a critical value of the initial population
$N_c = D q^2/\gamma$. For $N_0<N_c$ the total emission is very low being 
due to spontaneous processes that are only weakly amplified.  
On the contrary, for $N_0>N_c$ strong amplification occurs: 
the number of photons within the sample increases exponentially in time 
at a rate given by Eq.~(\ref{thr}), $r=\gamma N_0 - Dq^2=\gamma(N_0-N_c)$.
After the pulse has reached a maximum and the population is 
depleted, the emission decreases strongly. An estimate of the decay time 
of the pulse is given by solving the linearized equations around the
stationary state $N=0$, $I=0$. A straightforward calculation yields
that the long--time evolution is approximated by 
$N(x,t)= N_q(t)\sin(qx)$,
$I(x,t)= I_q(t)\sin(qx)$ where
\begin{eqnarray}
&& N_q(t) = A \exp(-\gamma t) \\
&& I_q(t) = A \frac{\gamma}{Dq^2-\gamma}\exp(-\gamma t) 
+ B \exp(- Dq^2 t), \nonumber 
\label{linear}
\end{eqnarray}
with $A, B$ being suitable time-independent amplitudes. 

The above results have been checked by comparing them with the 
numerical solution of (\ref{rateq}) obtained by simple integration methods 
of ordinary differential equations. In particular, we checked that both 
the rise and fall rates of the emission pulses (see the figures in the 
next section) are consistent with the expected values of $r$ and 
Eqs.~(\ref{linear}), respectively. 

\section{Monte Carlo simulations}
\label{s:mc}

In this Section we report the results of the simulation of the stochastic
model. Preliminary runs were performed to check that lasing thresholds exist
upon increasing of either the pumping parameter $N_0$ and the slab width $L$.  
The values are in agreement with the theoretical analysis presented above. 
In addition, checks of relations (\ref{expl}) and (\ref{lmed}) have 
been performed. 

As explained in Section \ref{s:model}, we monitored the outcoming
flux (per unit length) $\phi$ as function of time. The results are compared 
with the corresponding quantity evaluated from Eqs.~(\ref{rateq}). 
In this case, $\phi$ is defined from the discrete continuity equation to be 
\begin{equation}
\phi \;=\; \frac{D}{2}\Big[I(1,t) + I(L,t)\Big]\quad. 
\label{flux}
\end{equation}
The factor 2 comes from taking into account the contribution from 
the two boundaries $x=0,L$ of the lattice.
 
\begin{figure}[ht]
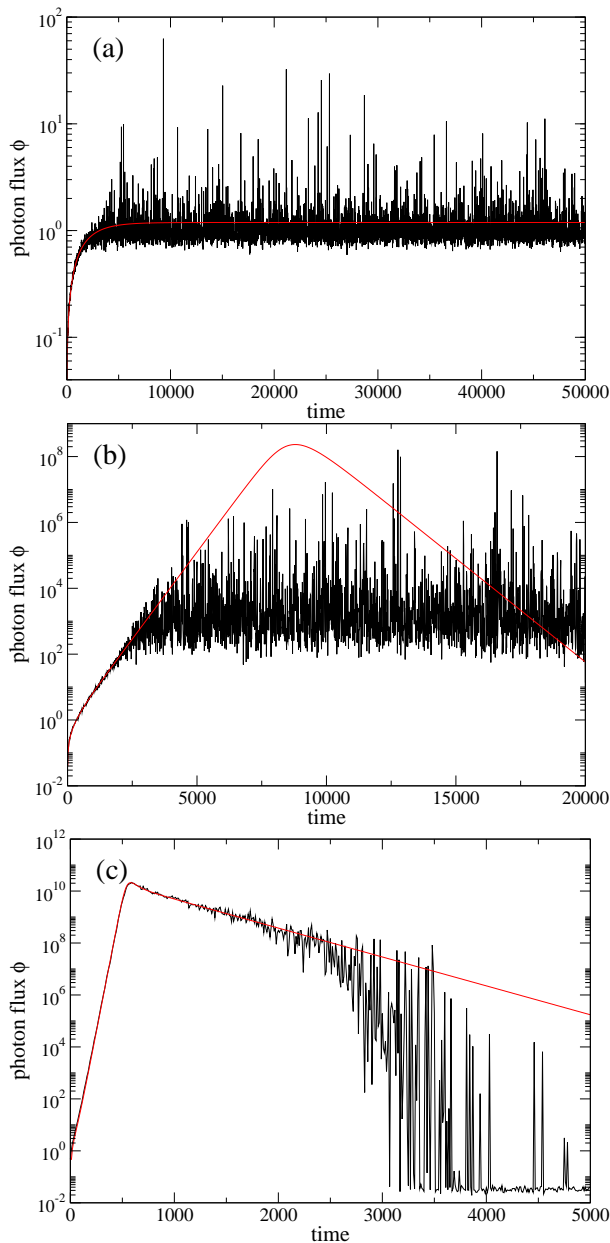

\begin{center}
\includegraphics[width=8cm,clip]{fig2a.eps}
\includegraphics[width=8cm,clip]{fig2b.eps}
\includegraphics[width=8cm,clip]{fig2c.eps}
\caption{(Color online) The photon flux (per unit length) as a function 
of time for a single shot, for $N_0=2 \times 10^9 $ 
(a); $N_0=5 \times 10^9 $ (b) and $N_0=50 \times 10^9 $ (c). 
Smooth red lines are the mean-field results, evaluated inserting the 
solutions of Eqs.~(\ref{rateq}) into Eq.~(\ref{flux}). For both curves, 
data have been binned over consecutive time windows of duration $T_W=10$. 
Note the difference in the vertical--axis scales.}
\label{f:pulse}
\end{center}
\end{figure}
The results of Monte Carlo simulation for a lattice with $L=30$, $R=20$ (18000
sites) and $\gamma=10^{-12}$ (yielding $N_c=2.5673 \times 10^9$) are reported in
Fig.~\ref{f:pulse}. The three chosen values of $N_0$ are representative of the
three relevant statistical regions depicted in Fig.~\ref{f:sog}: they correspond 
to $\ell_G=500$ (Subthreshold L\'evy), $\ell_G=200$ (Suprathreshold L\'evy) and
$\ell_G=20$ (Gaussian), respectively (see the triangles in Fig.~\ref{f:sog}).
In the first two cases, the total emission is highly irregular with huge 
deviations from the expected mean--field behavior. Above the lasing threshold 
(Fig.~\ref{f:sog}b) single events (``lucky photons") may carry values of 
$n_i$ up to $10^{10}$. The resulting time-series are quite sensitive to 
initialization of the random number generator used in the simulation.
On the contrary, in the Gaussian case (Fig.~\ref{f:pulse}c) the pulse
is pretty smooth and reproducible, except perhaps for its tails 
that, however, have a much smaller relative intensity.

\begin{figure}[ht]
\begin{center}
\includegraphics[width=8cm,clip]{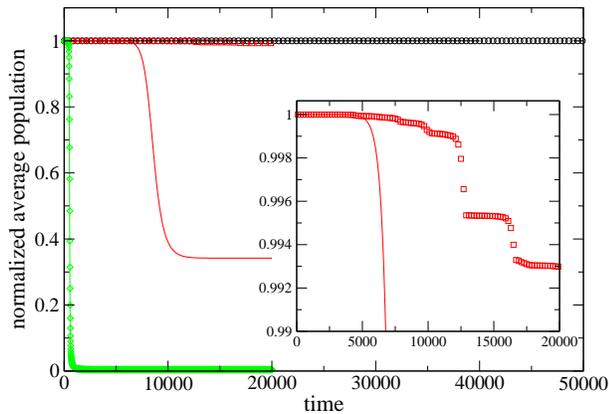}
\caption{(Color online) The normalized volume-averaged atomic population
as a function of time for a single shot and for the same values 
of $N_0$ as in Figs.~\ref{f:pulse}a-c (upper to lower curves
respectively). Solid lines are the mean-field 
results evaluated inserting the solutions of Eqs.~(\ref{rateq}) 
into (\ref{apop}). The inset shows a magnification of the middle curves 
(case $N_0=5 \times 10^9 $ of Fig.~\ref{f:pulse}b).
}
\label{f:pop}
\end{center}
\end{figure}

The evolution of the population $N$ displays similar features. We 
have chosen to monitor the volume--averaged population
\begin{equation}
\frac{1}{RL^2} \sum_{\textbf{r}}{N(\textbf{r},t)} 
\label{apop}
\end{equation}
normalized to its initial value for a better comparison.  
Fig.~\ref{f:pop} shows the corresponding time--series for the same 
runs of Fig.~\ref{f:pulse}. Again, large deviations from mean--field appear 
for the first two values of $N_0$. The inset shows that in correspondence 
with large--amplitude events the population abruptly decreases 
yielding a distinctive stepwise decay.

\begin{figure}[ht]
\begin{center}
\includegraphics[width=8cm,clip]{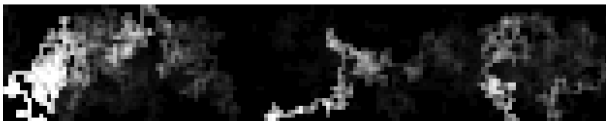}
\caption{A gray--scale plot of the atomic population distribution
along a portion of the lattice 
for $t=10^5$, $N_0=5 \times 10^9 $. White regions correspond
to small values of $N$.
}
\label{f:snap}
\end{center}
\end{figure}

The non-smooth time decay is accompanied by irregular evolution in space. 
Indeed, a snapshot of $N(\textbf{r},t)$ reveals a highly inhomogeneous 
profile (see Fig.~\ref{f:snap}). Light regions are traces of high--energy 
events that locally deplete the population before exiting the sample. 

For the Gaussian case, Fig.~\ref{f:pulse}c,  similar 
considerations as those made for the corresponding pulse apply. Note 
that now the population level decays extremely fast. It reaches 10\% of 
its initial value at $t\approx 600$ which is only twice the
average residence time within the sample. This means that photons emitted 
after a few hundreds time steps have hardly any chance to be significantly 
amplified (i.e. $\ell_G$ has become too large). 

\begin{figure}[ht]
\begin{center}
\includegraphics[width=8cm,clip]{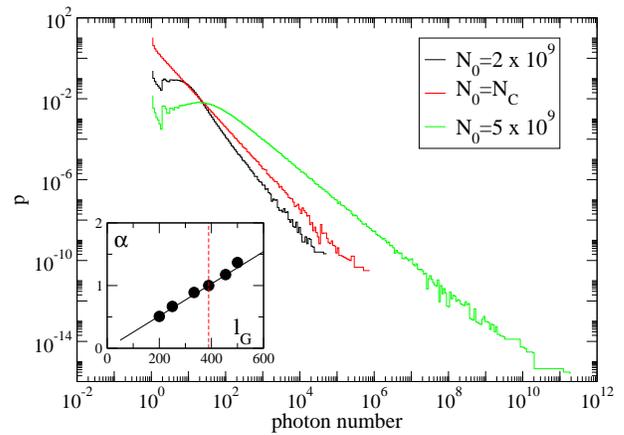}
\caption{(Color online) Histogram $p(n_{out})$ of the emitted 
photon numbers $n_{out}$
for the two values of $N_0$ of Figs.~\ref{f:pulse}a and b (lower
and upper curves) and $N_0=N_c$ (middle). 
This middle curve has been vertically shifted for clarity by roughly a factor
5.
The inset reports the values of $\alpha$ obtained by a power-law fit of the
histograms as a function of the gain length $\ell_G$ as defined by
Eq.~(\ref{lg}). The thin solid line is the theoretical curve as computed
by Eqs.~ (\ref{lmed}) and (\ref{pow}). The dashed vertical line 
marks the lasing threshold.}
\label{f:isto}
\end{center}
\end{figure}

To check directly the validity of the power--law distribution (\ref{pow}) we
computed the histograms of the photon number $n_{out}$ for each  and every
collected event during the whole simulation run (i.e. the  same time range as in
Fig.~\ref{f:pulse}).  The result are given in Fig.~\ref{f:isto} for three values
of $N_0$ for  which the L\'evy distribution (\ref{pow}) is expected to occur.  A
clear power-law tail extending over several decades is observed.  Note that the
middle curve correspond to the threshold value $N_0=N_c$  for which we expect
$\alpha=1$. Remarkably,  the values of the exponents measured by fitting the
data are in excellent  agreement with the definition of $\alpha$ (see inset of 
Fig.~\ref{f:isto}). As predicted, no meaningful value smaller than $\alpha=1/2$
is obtained  from the data. It should be noted that we are dealing with a
non--stationary process and the results may thus, in principle, depend on the
observation time. To check for a possible time-dependence of the statistics, 
we considered a four-times longer 
simulation for the case of Fig.~\ref{f:pulse}b and divide  the resulting time
series in four consecutive parts. Each of the  resulting histograms are almost
indistinguishable confirming that  the underlying process is almost stationary
at least on this time scale.

Finally, to further elucidate the differences between the two types of
statistical regimes, we performed a series of simulations increasing the number
of  lattice sites. For comparison, we kept $L=30$ fixed and increased the aspect
ratio $R$ up to a factor 4. In this way, we increased the number of walkers
accordingly. For the Gaussian case, we did observe the expected reduction of
fluctuations around the mean--field solution. On the contrary, the wild
fluctuations of the L\`evy case were hardly affected. This is a further
confirmation of the scenario discussed in Section \ref{s:gen}.

\section{Discussion}

Based on heuristic arguments, we have shown in Section \ref{s:gen} that, 
depending on the value of the dimensionless parameter ${D \Lambda \ell_G}/v$,  
the fluctuations in the emission of a random laser subject to short pump pulses
can be drastically different. In a parameter region extending both  above and
below threshold, the intensity fluctuations follow a L\'evy distribution thus
displaying wild fluctuations and huge differences in the emission from pulse to
pulse.  In the suprathreshold case, such features have been indeed observed in
experiments \cite{Sharma}. Some highly irreproducible emission with 
lack of self-averaging and very irregular behaviour has been also 
detected \cite{unpub}.

The exponent $\alpha$ of the L\'evy distribution can be tuned upon
changing the pumping level but it must be somehow bounded from below ($\alpha
\gtrsim 1/2$) as a further crossover to a Gaussian statistics is attained.
Indeed, far above threshold, when the gain length is very small, a
large and fast depletion of population occurs (saturation). This hinders the
possibility of huge amplification of individual events. In this case all photons
behave in a statistically similar way. As a consequence, the statistics is
Gaussian and a mean--field description applies again.

The above considerations have been substantiated by comparison with a simple
stochastic model. It includes population dynamics in a self--consistent manner.
In the L\'evy regions, the simulation data strongly depart from the predictions
of the mean--field approximation due to the overwhelming role of individual rare
events. As a consequence, the evolution of the population displays
abrupt changes in time and is highly inhomogeneous in space.

To conclude this general discussion we remark that the width of
the L\'evy region as defined by inequalities (\ref{bounds}) and depicted in
Fig.~\ref{f:sog} is of order $L^2$. Since in our simple model, $\ell_G$ is
inversely proportional  to the pump parameter (see Eq.~(\ref{lg})), 
the interval of $N_0$ values for which the L\'evy fluctuations occur shrinks as
$1/L^2$. Therefore, the larger the lattice, the closer to threshold one must be
to observe them.    

The existence of different statistical regimes, their crossovers and their 
dependence on various external parameters enriches the possible 
experimental scenarios. The emission statistics of random amplifying media 
has diverging moments in a \textit{finite} region of parameters extending 
across the threshold curve. Our theoretical work has shown that, depending 
on size, geometry, pumping protocols etc., the emission of random lasers 
may change considerably. This general conclusion should be a useful 
guidance in understanding past and future experiments on random amplifying
media. 

\section*{Acknowledgements}

We are indebted to R.\ Livi, S.\ Mujumdar, and A.\ Politi, for useful discussions and 
suggestions and to the {\it Centro interdipartimentale per lo Studio 
delle Dinamiche Complesse} (CSDC Universit\`a di  Firenze) for hospitality.
This work is part of the PRIN2004/5 projects {\it Transport properties of
classical and quantum systems} and {\it Silicon based photonic crystals} 
funded by MIUR-Italy, and was financially also supported by LENS
under EC contract RII3-CT-2003-506350, and by
the EU through Network of Excellence Phoremost (IST-2-511616-NOE).
G-LO thanks SGI for kind support.


\begin{thebibliography}{99}

\bibitem{book}
P.\ Sheng, {\em Introduction to Wave Scattering, Localization,
    and Mesoscopic Phenomena} (Academic Press, San Diego, 1995).

\bibitem{atom_loc}
H. Gimperlein et al., Phys.\ Rev.\ Lett.\ {\bf 95}, 170401 (2005);
D. Clement et al., Phys.\ Rev.\ Lett.\ {\bf 95}, 170409 (2005);
C. Fort et al., Phys.\ Rev.\ Lett.\ {\bf 95}, 170410 (2005).

\bibitem{Letokhov}
V.S.\ Letokhov, Zh.\ \'{E}ksp.\
Teor.\ Fiz.\ {\bf 53}, 1442 (1967)
[Sov.\ Phys.\ JETP {\bf 26}, 835 (1968)].

\bibitem{markushev}
V.M.\ Markushev, V.F.\ Zolin, Ch.M.\ Briskina,
Zh.\ Prikl.\ Spektrosk.\ {\bf 45}, 847 (1986);

\bibitem{migus93} C.\ Gouedard, et al.,
J.\ Opt.\ Soc.\ Am.\ B {\bf 10}, 2358 (1993).

\bibitem{lawandy94andsha94}
N.M.\ Lawandy, et al., Nature (London) {\bf 368},
436 (1994); W.L.\ Sha, C.H.\ Liu, and R.R.\ Alfano,
Opt.\ Lett.\ {\bf 19}, 1922, (1994).

\bibitem{bahoura02}M.\ Bahoura,
K.J.\ Morris, and M.A.\ Noginov, Opt.\ Comm.\ {\bf 201}, 405 (2002).

\bibitem{wiersma01nature} D.S.\ Wiersma and S.\ Cavalieri, Nature {\bf 414}, 708 (2001).

\bibitem{wiersma96} D.S.\ Wiersma and A.\ Lagendijk,
Phys.\ Rev.\ E {\bf 54}, 4256 (1996); {\it Light in strongly scattering and amplifying random systems},
D.S. Wiersma (PhD thesis, Univ.\ of Amsterdam, 1995).

\bibitem{cao99}
H.\ Cao, Y.\ G.\ Zhao, S.\ T.\ Ho , E.\ W.\ Seelig, Q.\ H.\ Wang,
and R.\ P.\ H.\ Chang,  Phys.\ Rev.\ Lett.\ {\bf 82}, 2278 (1999).

\bibitem{cao00b} H.\ Cao, J.\ Y.\ Xu, S.-H.\ Chang and S.\ T.\ Ho,
Phys.\ Rev.\ E {\bf 61}, 1985 (2000).

\bibitem{vardeny_overview}
R.C.\ Polson, M.E.\ Raikh, and Z.V.\ Vardeny,
IEEE J.\ Sel.\ Topics in Q.\ Elec.\ {\bf 9}, 120 (2003).

\bibitem{pradhan94}
P.\ Pradhan and N.\ Kumar, Phys.\ Rev.\ B {\bf 50}, 9644 (1994).

\bibitem{genack_loc}
V.\ Milner and A.Z.\ Genack, 
Phys.\ Rev.\ Lett.\ {\bf 94}, 073901 (2005).

\bibitem{Sushil} S.\ Mujumdar, M.\ Ricci, R.\ Torre, and D.\ S.\ Wiersma
Phys.\ Rev.\ Lett. {\bf  93}, 053903 (2004).

\bibitem{zyuzin94} A.\ Yu.\ Zyuzin, Europhys.\ Lett.\ {\bf 26}, 517 (1994).

\bibitem{john96} S.\ John and G.\ Pang,  Phys. Rev. A {\bf 54}, 3642 (1996).

\bibitem{Florescu} L. Florescu and S. John, Phys. Rev. Lett. 
\textbf{93} 013602 (2004); Phys. Rev. E \textbf{69} 046603 (2004)

\bibitem{altshuler91}B.L.\ Altshuler, V.E.\ Kravtsov, and I.V.\ Lerner, in
{\it Mesoscopic Phenomena in Solids}, edited by B.L.\ Altshuler, P.A.\ Lee, and R.A.\ Webb (North-
Holland, Amsterdam, 1991).

\bibitem{mirlin00} A.\ D.\ Mirlin, Phys.\ Rep.\ {\bf 326}, 259 (2000).

\bibitem{patra03}M.\ Patra, Phys.\ Rev.\ E {\bf 67}, 016603 (2003).

\bibitem{karpov93} V.\ G.\ Karpov, Phys.\ Rev.\ B {\bf 48}, 4325 (1993).

\bibitem{skipetrov04} S.\ E.\ Skipetrov, and B.\ A.\ van Tiggelen, Phys.\
Rev.\ Lett. {\bf 92}, 113901 (2004).

\bibitem{apalkov02} V.\ M.\ Apalkov, M.\ E.\ Raikh, and B.\ Shapiro, Phys.\ Rev.\ Lett. {\bf
89}, 126601 (2002)

\bibitem{Berger} G. A. Berger, M. Kempe and A. Z. Genack, Phys. Rev. E
\textbf{56} 6118 (1997)

\bibitem{Sushil2} S. Mujumdar, S. Cavalieri and D.S. 
Wiersma, J. Opt. Soc. Am. B {\bf 21}  201 (2004).

\bibitem{Jiang} X. Jiang and C.M. Soukoulis, Phys. Rev. Lett. 
\textbf{85} 70 (2000).

\bibitem{beenakker9800} C.W.J.\ Beenakker, 
Phys.\ Rev.\ Lett.\ {\bf 81}, 1829 (1998).

\bibitem{anglos} D.\ Anglos et al., 
J.\ Opt.\ Soc.\ Am.\ B {\bf 21}, 208 (2004).

\bibitem{vandermolen} K.\ van der Molen, A.P.\ Mosk, and 
A.\ Lagendijk, Phys.\ Rev.\ A {\bf 74}, 053808 (2006).

\bibitem{Sharma} D. Sharma, H. Ramachandran and N. Kumar, 
Fluct. Noise Lett. \textbf{6} L95 (2006).

\bibitem{Redner} S. Redner, \textit{A Guide to First-passage Processes},
(Cambridge University Press, Cambridge, 2001). 

\bibitem{Levy} J.P. Bouchaud and A. Georges, Phys. Rep. \textbf{195}, 
127 (1990).

\bibitem{Carma} H.J. Carmichael, \textit{Statistical Methods in 
Quantum Optics 1} (Springer-Verlag, Berlin, 1999).

\bibitem{note} In our units this correspond to the condition 
$\gamma N n \ll 1$. For the parameters in the simulations 
$\gamma N \sim 10^{-3}$, i.e. the condition may be violated for 
short times. On the other hand this initial regime is irrelevant
for the effects we are interested in.

\bibitem{unpub} S. Mujumdar, V. Tuerck and D.S. 
Wiersma, unpublished.

\end{thebibliography}
\end{document}